**Uncovering the influence of common nonmetallic impurities on the stability and strength of a Σ5 (310) grain boundary in Cu**

Zhifeng Huang[a, b], Fei Chen[a, *], Qiang Shen[a], Lianmeng Zhang[a], Timothy J. Rupert[b, c, *]

[a] State Key Lab of Advanced Technology for Materials Synthesis and Processing, Wuhan University of Technology, Wuhan 430070, China

[b] Department of Mechanical and Aerospace Engineering, University of California, Irvine, CA 92697, USA

[c] Department of Chemical Engineering and Materials Science, University of California, Irvine, CA 92697, USA

* Corresponding author: chenfei027@whut.edu.cn (F. Chen), trupert@uci.edu (T.J. Rupert)

Impurities are often driven to segregate to grain boundaries, which can significantly alter a material's thermal stability and mechanical behavior. To provide a comprehensive picture of this issue, the influence of a wide variety of common nonmetallic impurities (H, B, C, N, O, Si, P and S) incorporated during service or materials processing are studied using first-principles simulations, with a focus on identifying changes to the energetics and mechanical strength of a Cu Σ5 (310) grain boundary. Changes to the grain boundary energy are found to be closely correlated with the covalent radii of the impurities and the volumetric deformations of polyhedra at the interface. The strengthening energies of each impurity are evaluated as a function of covalent radius and electronegativity, followed by first-principles-based tensile tests on selected impurities. The strengthening of a B-doped grain boundary comes from an enhancement of the charge density among the adjacent Cu atoms, which improves the connection between the two grains. Alternatively, the detrimental effect of O results from the reduction of interactions between the Cu atoms. This work deepens the understanding of the possible beneficial and harmful effects of impurities on grain boundaries, providing a guide for materials processing studies.

## 1. Introduction

Grain boundaries are ubiquitous defects in the vast majority of engineering materials, where they play an important role in governing a range of mechanical, functional and kinetic properties [1-6]. These properties can be affected by local chemistry, such as the common nonmetallic impurities such as H [7-10], B [11-14], C [14-16], N [16, 17], O [16-18], Si [19, 20], P [20, 21] and S [20, 22, 23] that are abundant in service environments and conventional processing routes. For example, H embrittlement is the widespread problem and can result in sudden catastrophic failure of metallic components [9]. Alternatively, adding an appropriate amount of C and B to steels can dramatically increase their hardness and toughness. This improvement comes as a result of a transition from grain boundary nucleation of $Fe_{23}(CB)_6$ to intragranular nucleation of this phase [14]. In addition, impurities such as C, O, and H are often incorporated into materials produced through mechanical alloying in the presence of a process control agent [24-26]. For example, the dissolution of C was found to have a positive effect on the thermal stability of an Fe–Zr nanostructured alloy created by ball milling [26]. Although a signficant amount of experimental work has shown the importance of these nonmetallic impurities, the dominant mechanisms responsible for these effects must be studied on an atomic and electronic level.

First-principles simulations [27] are ideal tools for the study of material defects at the atomistic and electronic levels, having been used previously to investigate grain boundary. Huber et al. [28] found that oversized dopants (Mg and Pb) showed a trend of stronger segregation to the site with a larger Voronoi volume in an Al grain boundary. Wu et al. [29] demonstrated that different metallic dopants prefer different substitutional sites at a given grain boundary, due to their different atomic sizes. For grain boundaries with nonmetallic impurities, previous work has shown that impurities favor grain boundary interstitial sites rather than substitutional sites at the interfaces [11, 19, 20, 23, 30]. Zhang et al. [18] showed that the preferred segregation sites for small nonmetallic impurities (C, N and O) and large impurities (Si and S) are different at a Σ3 (111) grain boundary in V. It is worth noting that Σ3 (111) grain boundaries have excess volume in body centered cubic metals, which are not close packed. Furthermore, Du et al. [31] found that the segregation ability of H within different Fe grain boundaries strongly depended on the local coordination of the corresponding interstitial site. Zhang et al. [32] and Hatcher et al. [33] found that the solution energy of He and C in Fe grain boundaries decreased with an increase of the Voronoi volume of the occupied site. Alternatively, Zhou et al. [34, 35] provided a quantitative evaluation of H adsorption energies as a function of the volumetric deformation of polyhedra at grain boundary sites in a variety of face-centered cubic metals. These results show that the segregation ability of H is closely related to the type of polyhedra at grain boundary. Unfortunately, such a relationship has not been investigated in detail for other common nonmetallic impurities. In this work, we explore a comprehensive set of nonmetallic impurities and look for important trends based on recently developed grain boundary metrics.

The impact of impurities on the mechanical strength of interfaces has also been a topic of many investigations [11, 14, 17, 18, 30], which shows that certain impurities can drastically improve the mechanical strength while others cause a detrimental effect. For example, B has been reported to improve the cohesion at grain boundaries in Cu [11], Mo [17] and Fe [36]. Some studies suggest that impurities can form covalent bonds with the host atoms to strengthen interfacial cohesions while others form isotropic polar bonds that cause interfacial embrittlement [17, 18, 21]. Geng et al. [12] showed that the condition for one impurity to be an enhancer was that the atomic size of the impurity must allow it to fit well into the grain boundaries, being neither too small nor too large. The examples given above suggest that atomic size, the details of atomic bonding between impurities and the host atoms, and the type of polyhedra at the grain boundary are candidates for controlling the behavior of nonmetallic impurities and their segregation to interfaces. To narrow the search and identify one underlying cause, a systematic exploration of these various factors is needed, for a full set of possible nonmetallic impurities.

In addition to understanding how such impurities affect the mechanical response of grain boundaries, it is also important to understand how such doping alters the thermal stability of interfaces. This is especially important for interface-dominated materials, such as nanocrystalline metals [37-39] and nanolaminates [40, 41]. From the thermodynamic mechanism, the lack of thermal stability of these materials is often attributed to the high grain boundary fraction, which gives a large driving force $P$ for coarsening that can sometimes result in the grain growth even at room temperature in pure nanocrystalline metals [42, 43]. The driving force is proportional to the grain boundary energy $\gamma$, which is a linear function of temperature as shown by the Gibbs adsorption isotherm [44]. Combining the Gibbs adsorption equation and Mclean's grain boundary segregation model, Liu and Kirchheim [45] predicted that grain boundary energy could be reduced, possibly even to zero to make a fully stable material, through solute segregation. Building on this idea, a number of experimental and theoretical research studies have shown that the thermal stability of nanostructured metals could be significantly improved by metallic dopant segregation to the interfaces [39, 43, 45-47]. However, the impact of nonmetallic impurities on thermal stability has undergone much less study. Chen et al. [48] and Li et al. [49] showed that the segregation of C at the grain boundaries could improve the stability of nanocrystalline Fe alloys, and Koch et al. [43] found that P could stabilize nanocrystalline Ni. However, a systematic investigation of the effect of nonmetallic impurities on thermal stability is lacking. While nanocrystalline metals are quoted here as a motivating class of materials, due to the high density of grain boundaries and their importance for determining the properties of these materials, the effects of impurities on grain boundary energy and strength are important for all polycrystalline metals. Because these common impurities can be incorporated during either processing or service, the topic of impurity segregation to grain boundaries should be broadly of interest to metallurgists.

In this paper, we perform a systematic exploration of the impact of common nonmetallic impurities (H, B, C, N, O, Si, P and S) on the energetics and mechanical strength of a Σ5 (310) grain boundary in Cu using first-principles calculations. The Σ5 (310) was chosen as a model boundary due to the presence of three different structural polyhedral at the interface that act as potential segregation sites. This variety is more than some other common boundaries, such as the Σ5 (210) which only has two boundary interstitial sites, one of which is extremely small and unlikely to accept larger impurities. First, we explore correlations between energetic variables such as the segregation energy, grain boundary energy and strengthening energy with major descriptive features of the doping elements such as the covalent radius and the electronegativity of impurities, as a function of the type of polyhedra at the interface. Grain boundary energy is found to decrease as the covalent radius of the impurities and the volumetric deformation of interfacial polyhedra increases. For impurities within a given period of the periodic table, the strengthening energy exhibits a positive correlation with the electronegativity of impurities, while demonstrating a negative correlation with the atomic radius. Next, we perform first-principles-based tensile tests to uncover the interactions between selected impurities and the host Cu atoms, taking care to bond reconstructions along the fractured surface during the tensile straining since prior work has left out such behavior and used only a Rice-Wang model [50] to investigate the strengthening effect of impurities. The opposite effects of B and O on grain boundary strength come from the fact that B prefers a site where it enhances the interactions among the closest Cu atoms to improve the connection between the two grains, while O segregates into another site that reduces these interactions between the Cu atoms. With the detailed atomistic and electronic information provided by these first-principles calculations, this work illuminates the possible beneficial and harmful effects of impurities on grain boundary properties, which must be considered during materials processing.

## 2. Computational methods and details

All calculations were performed using density functional theory (DFT) implemented in the Vienna *ab-initio* simulation package (VASP) using the projector augmented wave approach [51] and the Perdew-Burke-Ernzerhof exchange-correlation generalized gradient approximation (GGA) functional [52, 53]. A plane-wave cutoff energy of 350 eV, convergence energy of $10^{-5}$ eV/atom, and convergence atomic force of 0.01 eV/Å were used for all calculations. First, an $8 \times 8 \times 8$ *k*- point mesh was used to find the initial Cu unit cell. The calculated equilibrium lattice parameter of fully relaxed face centered cubic Cu unit cell is $a$ = 3.634 Å under our convergence criteria, which is consistent with previous DFT-GGA calculations from Haas et al. (3.632 Å) [54]. Then the grain boundary model was generated by doubling the periodic lattice parameters in the [001] and [1-30] directions, choosing 27 atomic layers and a 12 Å vacuum layer in the [310] direction based on the Cu unit cell. Thus, the grain boundary specimens have dimensions of $7.267 \times 11.490 \times 28.086$ Å$^3$ and are comprised of 112 atoms. The bulk model and free surface specimens have the exact same dimensions as the grain boundary model, but contain 112 and 60 atoms, respectively. Considering the large number of models used in this work, covering a range of impurities, possible segregation sites, and also first-principles tensile tests, we reduced the $k$-point meshes to $3 \times 2 \times 1$ for all of the grain boundary, bulk, and free surface models to balance the accuracy and efficiency of calculation. The atoms in bulk, grain boundary, and free surface samples were fully relaxed during the process of structural relaxations, except for the $z$ coordinate of atoms on the outermost layers in each model, which were fixed. The dimensions of the bulk sample, grain boundary, and free surface were fixed during structural optimizations to keep the cross-section area ($S$) the same for the grain boundary energy calculations, as shown in Eq. (3). From Fig. 1, there are two possible interstitial sites in the Cu lattice: the octahedron (OCT) and tetrahedron (TET) sites, while three possible interstitial sites at the $\Sigma5$ (310) grain boundary: the pentagonal bipyramid (PBP), bitetrahedron (BTE), and cap trigonal prism (CTP) sites [55]. The grain boundary has four types of possible substitutional sites for impurities at the interface, labeled as 1-4 in Fig. 1(b), and two different planes in the [001] direction, as marked as plane 1 and plane 2 in Fig. 1(c) [34]. The calculated excess volume of the Cu Σ5 (310) grain boundary is 0.30 Å, which is consistent with previous calculations of 0.28 Å from a DFT study using a Local-Density Approximations (LDA) functional [11].

The ability of an impurity, X, to segregate to the grain boundary can be characterized by the segregation energy ($E_{seg}$) [56] given by Eq. (2):

$$E_{seg} = (E_{GB+X} - E_{GB}) - (E_{bulk+X} - E_{bulk}) \quad (2)$$

where $E_{GB}$, $E_{GB+X}$, $E_{bulk}$ and $E_{bulk+X}$ are the total energy of the undoped grain boundary, the grain boundary with the impurity X, the undoped bulk Cu, and the bulk sample with the impurity X, respectively. $E_{bulk+X}$ is determined as the lowest energy found when inserting the impurity in the two possible polyhedra in the crystalline Cu. A negative segregation energy indicates that an impurity segregating into the site is energetically favorable [57]. The preferred segregation site for an impurity is then found by comparing the segregation energies of X in different substitutional and interstitial sites at the interface. A more physically meaningful metric for thermodynamic stability is the grain boundary energy ($\gamma$), which can be defined as the difference between the energy of a sample with a grain boundary and that of a bulk crystal, normalized by the grain boundary area [58-60]:

$$\gamma = \frac{E_{GB/GB+X} - E_{bulk/bulk+X}}{S} \quad (3)$$

The calculated grain boundary energy of the clean Cu Σ5 (310) grain boundary is 0.88 J/m$^2$, which is identical to a previous calculation of 0.88 J/m$^2$ from DFT-GGA [61].

A preliminary prediction of the impact of impurities on grain boundary strength can be measured by the strengthening energy ($E_{str}$) based on Rice-Wang model:

$$E_{str} = (E_{GB+X} - E_{GB}) - (E_{FS+X} - E_{FS}) \quad (4)$$

where $E_{FS}$ and $E_{FS+X}$ stand for the total energies of the undoped Cu free surface and free surface with an impurity atom. The site choices of X for the free surface model are the same as in the grain boundary model. A negative value of strengthening energy means that the impurity will enhance the grain boundary strength, while a positive value suggests a detrimental effect on strength.

There are two common types of stretching schemes found in the literature for first-principles-based tensile tests. In the first scheme, the atomic configuration in each strain step is found by uniform elongation of the fully relaxed configuration of the preceding step [62-64]. In this scheme, the model should break by itself at the weakest point above a certain critical strain. In the second scheme, one fracture plane is set a priori and a pre-crack is introduced at this plane [22, 29, 64, 65]. Generally, the first scheme is very time-consuming, and the structural relaxation often becomes difficult to converge when the model has a large strain near the fracture point. Therefore, we choose to use the second scheme in this work.

The simulation cell used for the first-principles-based tensile tests is shown in Fig. 2(a), where the two potential fracture paths are denoted by the red and black lines (P1 and P2). Since these calculations are more computationally expensive, we limit ourselves to the clean grain boundary, a grain boundary with B at a CTP site, and a grain boundary with O at a PBP site. Noting that the grain boundary just contains one B or O atom, the grain boundary concentration of B or O is 4 at.% or 1.2 atoms/nm$^{-2}$. Previous work has shown that B prefers to stay as an isolated impurity (rather than a B-B dimer) at the interface when the grain boundary concentration is low [11], and isolated O atoms have also been observed in lightly deformed Ti with 0.1-0.3 wt.% O [66]. These previous observations indicate that isolated impurity atoms in the grain boundary can be the stable configuration, rather than compounds such as borides, oxides, or nitrides which can form at higher impurities levels. These two nonmetallic impurities are chosen because both elements are in the second period, but they prefer different sites at the interface and have opposite effects on the grain boundary strength, as shown in Section 3.2. A series of separation distances are inserted between the upper and lower crystal blocks at the fracture plane location. For each separation distance, two kinds of calculation are performed: (1) rigid, without relaxation of the atomic positions, and (2) relaxed, with full atomic relaxation except for the *z* axis of atoms on the outermost layers at the top and bottom. The separation energy ($E_{sep}$) can be obtained by Eq. (5):

$$E_{sep} = \frac{E_x - E_0}{S} \tag{5}$$

where $E_0$ is the energy of the grain boundary without separation, $E_x$ is the energy of the fracture boundary with a separation distance of $x$. The limit of the separation energy at infinite separation, defined as the fracture energy. The universal binding energy relation proposed by Rose et al. [67] is used to fit the separation energy of the rigid calculation following:

$$f(x) = E_{sep} - E_{sep}(1 + \frac{x}{\lambda})e^{(-x/\lambda)} \tag{6}$$

where $\lambda$ is the characteristic separation distance.

## 3. Results and discussion

### 3.1 Segregation, grain boundary, and strengthening energies

Fig. 3 presents the segregation energies of B, O, and Si impurities at different substitutional and interstitial sites at the interfaces. The result shows that the impurities all strongly prefer the interstitial sites as opposed to the substitutional sites at the interface, with most of the substitutional sites even giving a positive segregation energy, meaning the impurity would prefer to be at a substitutional site inside the bulk crystal. Lozovoi et al. [11] also found that B preferred to segregate to an interstitial site (only the CTP site was considered in this study) rather than the substitutional sites when studying a Cu Σ5 (310) grain boundary using a DFT-LDA method. However, the preferred site for O is the PBP site while B and Si prefer the CTP site, indicating that the preferred interstitial site is material-dependent. It is important to note that when B and Si are added to a BTE site and the structure relaxed, the impurity atoms will move to either the PBP or CTP sites, meaning that no stable energy can be calculated for the BTE site.

With a focus on the interstitial sites, we next move to investigate the full range of nonmetallic impurities. Fig. 4 shows the segregation, grain boundary and strengthening energies of different nonmetallic impurities at the various possible interstitial sites at the boundary. Fig. 4(a) shows that B, Si, P and S prefer the CTP site while H, C, N and O prefer the PBP site. This observation suggests that the impurities with a relatively large atomic size prefer the larger sites while those with a small atomic size prefer the smaller sites. This connection between relative atomic size and the size of the potential segregation interstitial site was also observed by Zhang et al. [18] for a Σ3 (111) boundary in V. Fig. 4(b) shows that the energies of all of the grain boundaries with impurities are lower than 0.88 $J/m^2$, indicating that the grain boundary energy has been reduced by impurity addition. From Fig. 4(a) and (b), it is clear that the changes to grain boundary energy with doping mirror the segregation energies, an observation that is not wholly surprising since the metrics rely on many of the same energetic parameters. Fig. 4(c) shows that B strengthens the boundary (negative strengthening energy), C has almost no effect, and all other impurities act to weaken the grain boundary (positive strengthening energy). Lozovoi and Paxton [11] hypothesized that O and N might weaken the Cu Σ5 (310) grain boundary, which is confirmed by our findings. For all impurities except from H, the strengthening energies of impurities at the CTP site are lower than at either the PBP or BTE sites.

Generally, it is advantageous to improve thermodynamic stability of a microstructure and to reduce the tendency for grain boundary fracture, corresponding to reductions in the energy values presented in Fig. 4. All of the nonmetallic impurities increase thermodynamic stability by decreasing grain boundary energy, but this often comes at the cost of an associated embrittling effect. B is a notable exception that improves both stability and strength, meaning it can be a useful additive during processing. The fact that C has almost no effect on grain boundary strength when located in the preferred PBP site is also of note. As mentioned in the introduction, many process control agents for mechanical alloying are made of C-based molecules, leading to the incorporation of C during ball milling. However, since this addition improves stability but does not have a detrimental effect on mechanical properties, one does not need to take drastic steps to reduce C contamination. Evaluation of the rest of the impurities then depends on the relative important of stability and interfacial strength for a given application. For example, if thermal stability is the main consideration, Si would be a useful additive because the dramatic reduction in grain boundary energy might be worth the very small embrittling effect when located in the preferred CTP site.

3.2 Factors controlling segregation and strengthening

With the overall effect of nonmetallic impurities on the stability and strength of a grain boundary shown by our first-principles calculations of segregation, grain boundary, and strengthening energies, we turn our attention to the factors controlling such behavior. The data points presented in Fig. 4 show that there is a large variation of segregation and strengthening energies for a given impurity when different sites are considered. Inspired by previous work that looked at the correlation between the adsorption ability of H and the volumetric deformation of the grain boundary polyhedra [34, 35], we consider the volumetric deformation as a useful quantity to consider, because it incorporates the impurity size while also normalizing by the site geometry and size. The volumetric deformation of a given polyhedron can be defined as $V_p = (V_P^X - V_P^0)/V_P^0$ [34, 35], where $V_P^0$ is the Voronoi volume of the pristine polyhedron and $V_P^X$ is the Voronoi volume of the polyhedron after the impurity X is added. The Voronoi volumes were analyzed via the open-source software VESTA [68]. The pristine volumes of the PBP, CTP and BTE are 12.508, 22.033 and 4.742 Å$^3$, respectively. We also consider the covalent radius and electronegativity of each impurity because these parameters are related to the size and chemical effects, respectively, associated with a given impurity [18, 69].

Fig. 5 presents all of the grain boundary energy data plotted against the covalent radius and electronegativity of the impurities, as well as the volumetric deformation of the doped polyhedra. Grain boundary energy decreases with increasing covalent radius of the impurities and the volumetric deformation of polyhedra. Linear fits are applied to the data for different site types in Fig. 5(c) to guide the eye, but we do not imply that there is a physical meaning to this fitting form. In contrast, there is no clear trend connecting grain boundary energy and the electronegativity of the nonmetallic impurities. Grain boundary energy is related to both covalent radius and volumetric deformation due to the connection of these parameters to a reduction in grain boundary free volume. A major contributor to the energetic penalty for a grain boundary comes from the fact that atoms in these regions are not as densely packed as they are in the lattice. Uesugi and Hagashi [70] demonstrated this concept clearly using first-principles calculations of [110] symmetric tilt grain boundaries in pure Al. These authors found that grain boundary energy increases linearly as the excess free volume of the grain boundary increases. Coming back to our work, Fig. 5(a) shows that impurities with larger covalent radii more efficiently fill free space in the boundary and reduce free volume. The volumetric deformation of the grain boundary polyhedral represents another view of this effect, as it incorporates the elastic distortion of the interstitial sites. This metric is biased by the site that is being filled, which explains why Fig. 5(c) varies as the interstitial site is changed (i.e., a correlation is found within a set of either PBP, CTP, or BTE sites). Although these findings are new for a large collection of common nonmetallic impurities, they are consistent with prior work on metallic grain boundary dopants. For example, Millett et al. [71] used molecular dynamics with a Lennard-Jones potential to parametrically study the effect of the atomic radius of dopants on grain boundary energy in face centered cubic metals, finding that the boundary energy decreased with increasing atomic radius misfit and interfacial coverage of dopants at the grain boundary. Zhou et al. [34] found that the adsorption energy of H in face-centered cubic $\Sigma 5$ grain boundaries decreased linearly as the volumetric deformation of polyhedral increased. Our work here shows that such trends extend beyond just H, to a complete set of nonmetallic impurity dopants. Our results also show that the connection between boundary energy and volumetric site depends strongly on the type of site that is doped, with the most negative slope in Fig. 5(c) being found for CTP sites, followed by PBP and BTE, respectively. This indicates that the local coordination of polyhedral sites plays an important role in the selection of dopant sites.

Fig. 6 shows the strengthening energy of the different nonmetallic impurities, plotted as a function of the covalent radius, electronegativity, and the volumetric deformation of interfacial polyhedra. As a reminder, these data points are data obtained through a Rice-Wang calculation. The strengthening energy appears to increase as covalent radius decreases and as the electronegativity of the impurities increases. In Fig. 6(a) and (b), linear fits are again drawn to guide the eye, but different curves are presented for impurities from different periods in the periodic table, for reasons explained in the following paragraph. From Fig. 6(c), our results also show a general increasing trend of strengthening energy with increasing volumetric deformation of polyhedral, although the data is scattered.

To uncover the mechanisms behind these effects on the grain boundary strength, the change of charge density and bond length at the interface were investigated. Fig. 7 shows the differential charge density (charge density change from the doped to the undoped case), with impurities added to the CTP site and the image showing a projection along the (100) plane. These results indicate that the interactions between impurities and the host Cu atoms are predominantly localized at the interfaces, meaning any changes to the strengthening energy is closely linked with the local atomic structure. While the interactions between impurity atoms and the surrounding Cu differ, general trends can be made by considering the period where the impurity atom is located in the periodic table. For H, which is located in the first period, the H impurity traps electrons (shown by the strong red color at the doping site) from only one of the nearby Cu atoms. For the impurities in the second period (B, C, N, and O), all of the Cu atoms closest to the impurities lose electrons to the segregation site, with this effect becoming more pronounced with increasing atomic number. For atoms in the third period (Si, P, and S), the electrons are mainly gathered in between the impurities and the surrounding Cu atoms. From Fig. 7, we can see that all of the impurities interact with their closest Cu1 atom (Cu atom in site 1 in Fig. 1), providing motivation to investigate the density of states of Cu1 and impurities to analyze the bond properties. The density of states data for Cu1 appear as thin curves while the impurities appear as heavy or thick curves in Fig. 8. For H, the *s* band has little hybridization interaction with the main *d* band of Cu1, which is similar to the reported "*s*-like band" for H at grain boundaries in Mo [17] and Fe [72]. This indicates that the Cu1-H is a strongly polar bond. For all impurities besides H, the main interactions are reflected in the hybridizations between the *d* states of the Cu and *p* states of the impurities. As the atomic number of the impurities increases in each period, the main peaks of *p* bands of impurities move to lower energies and become narrower, resulting in the weaker hybridization between the *d* states of Cu and *p* states of impurities. This change is similar to a prior report for B, C, N and O in a Mo grain boundary [17], which showed that the transition of the bond between Cu and impurities transitions from covalent to ionic. Electronegativity can generally be used to evaluate the ability of an atom to attract electrons towards itself [73] and the difference between the electronegativity for atoms in the second period is the largest, explaining why we see the largest change in bonding character within the period two impurities from Fig.7 and Fig. 8. With these observations in mind, we return to the data presented in Fig. 6(b). Similar to Fig. 5(c), we fit based on possible doping site, but now also distinguish between impurities in different atomic periods and begin to see trends emerge. Within a given period, an increase in the electronegativity of the dopant increases the strengthening energy, leading to embrittlement.

To show this effect more clearly, we highlight how B and O, both elements in period two but with a large difference in electronegativity, lead to bond reconstruction at the boundary. Fig. 9 shows the bond length changes of the Cu–Cu bonds around the impurity in grain boundaries with B and O at different sites, in an effort to understand how doping affects the details of atomic structure. The change of bond length is defined as $(l_X - l_0)/l_0$, where $l_X$ is the length of Cu–Cu bond near the impurity X and $l_0$ is the corresponding Cu–Cu bond length in the undoped grain boundary. The bond length is the separation distance between the two Cu atoms. Fig. 9(a) shows that the changes of all the Cu–Cu bond lengths around B at the PBP site are significant larger than those around B at the CTP site. Similarly, Fig. 9(b) shows that the changes of Cu–Cu bond lengths near O at the BTE site are the highest. When comparing these observations to Fig. 6, it is clear that impurities that lead to large changes in the bond lengths of the surrounding Cu atoms have the most detrimental effect on the grain boundary strength.

### 3.3 First-principles-based tensile tests

First, we calculated the fracture energies of the clean Cu Σ5 (310) grain boundary along different fracture paths, as shown in Fig. 2(a), to determine the fracture plane to be used for the relaxed first-principles-based tensile tests. The fracture energy along P1 is 2.266 $J/m^2$, which is less than the measured value of 2.678 $J/m^2$ along P2. This means that P1 is the preferred fracture path for the Cu Σ5 (310) grain boundary and the separated two free surfaces after fracture are shown in Fig. 2(b). Since our model just contains one B or O atom at the interface, the preferred fracture path of a grain boundary with either B or O can be determined by comparing the fracture energies of the systems with B or O at the upper or lower surfaces. The calculated fracture energies of both B (2.371 $J/m^2$) and O (2.097 $J/m^2$) at the lower surface are lower than the energies when B (2.430 $J/m^2$) and O (2.229 $J/m^2$) are at the upper surface, indicating that the nonmetallic dopants prefer to be located at the lower surface when fracture commences.

Fig. 10(a) and (b) show the rigid and relaxed separation energies of the clean grain boundary, the grain boundary with B at the CTP site, and the grain boundary with O at the PBP site, as well as the universal binding energy relation fit curves using the Rose formulation for the data in part (a). As shown in Fig. 10(a), the rigid separation energies follow the universal binding energy relation fit curves very well, which increases rapidly at the beginning, before slowing and eventually reaching an asymptote. The relaxed separation energy curves shown in Fig. 10(b) can be divided into three distinctive regions. For small separation distances, the separation initially introduced into the grain boundary is healed by atomic reconstruction during the relaxed geometry optimization, resulting in the reconnection between the two pre-fracture surfaces and making this region elastic. The separation energy as a function of separation follows Hooke's law in this range and is parabolic. The second region can be called the plastic stage, as the pre-crack can no longer heal and the separation energy continues to increase until the sample is fully broken and two new surfaces are created. The separation energy at the end of this stage is the ultimate strength. From the inset to Fig. 10(b), it is clear that the ultimate strength of the clean grain boundary is lower than that of the grain boundary with B at the CTP site yet higher than the strength of the grain boundary with O at the PBP site. In the third and final deformation stage, the separation energy increases slowly until it asymptotes as the remaining long-range interaction forces between the two fracture surfaces disappear. The relaxed separation energy curves at the third stage follows the universal binding energy relation and thus the fracture energy can be extracted.

The fracture energies calculated from the rigid and relaxed first-principles-based tensile test are included in Table 1. Both rigid and relaxed calculations for fracture energy show that the grain boundary with B at the CTP site is strongest, followed by the clean grain boundary, which is in turn followed by the grain boundary with O at the PBP site. These calculations are consistent with the ordering of the strengthening energies calculated from Rice-Wang model, but do contain more accuracy and nuance. In order to clarify the underlying connection of Rice-Wang theory and first-principles-based tensile tests, the following equations can be used [29]:

$$E_{frac}^{GB} = E_{\infty}^{GB} - E_{0}^{GB} = (E_{FS1} + E_{FS2}) - E_{GB} \quad (7)$$

$$E_{frac}^{GB+X}S = E_{\infty}^{GB+X} - E_{0}^{GB+X} = (E_{FS1+X} + E_{FS2}) - E_{GB+X} \quad (8)$$

Thus, the difference between the fracture energies of a clean and doped grain boundary with impurity X is:

$$\left(E_{frac}^{GB} - E_{frac}^{GB+X}\right)S = \left(E_{GB+X} - E_{GB}\right) - \left(E_{FS1+X} - E_{FS1}\right) \tag{9}$$

Eq. (9) is equivalent to Eq. (4) for the calculation of strengthening energy of grain boundary with impurity X, meaning the first-principles-based tensile tests can also be used to isolate the strengthening energy of different impurity dopants, as presented in Table 1. The strengthening energies of grain boundaries with B and O from the relaxed first-principles-based tensile tests are very close to the values of -0.547 eV and 0.888 eV given by the Rice-Wang model, with the relaxed calculations (-0.547 eV and 0.881 eV, respectively) closer than the rigid calculations (-0.417 eV and 1.115 eV, respectively). This consistency suggests that, while first-principles-based tensile simulations allow for more nuanced and detailed analysis, the preliminary Rice-Wang calculations offer reliable predictions of strengthening energies.

The tensile strength of the various samples can be taken as the maximum tensile stress, with the results presented in Fig. 10(c) and (d) and listed in Table 1. Both the rigid and relaxed calculations show that the tensile strength of the grain boundary with B at the CTP site is highest, followed by the clean grain boundary, which is then followed by the grain boundary with O at the PBP site. The different values of theoretical tensile strengths result from the different accuracies of rigid and relaxed first-principles-based tensile tests. From these simulations, we can see that the effects of B and O are significantly different, which will be explained by charge density distributions and bond reconstruction between the two fracture surfaces during straining in the discussion that follows.

Fig. 11 shows the charge density distributions in plane 1 and plane 2 of the three specimens during the relaxed first-principles-based tensile tests. The separation is labeled at the bottom of each figure part. As the separation distances increase, the charge densities between Cu atoms along the grain boundary in plane 1 slowly decrease until reaching the ultimate strength at displacements of 0.16 nm for the clean Cu grain boundary, 0.19 nm for the grain boundary with B, and 0.175 nm for the grain boundary with O. Meanwhile, in plane 2, the charge densities decrease in intensity and begin to disappear at a separation distance of 0.1 nm for the clean and O doped boundaries, while strong interactions persist to the ultimate strength at a displacement of 0.19 nm for the grain boundary with B. After the ultimate strength point, the strong electronic interaction between the two grains has almost disappeared.

Fig. 12 shows the length of the interfacial bonds as a function of the separation distance for the three samples. First, close inspection shows that most of the bond lengths between Cu–Cu atoms that surround the B atom in the grain boundary are shorter than those in the clean boundary, while most of the Cu–Cu bonds around the O atom are longer than those in the clean grain boundary. The changes to the Cu–Cu bonds adjacent to the B and O atoms reflect that B can enhance while O reduces the connection of Cu atoms. Furthermore, the changes to the bond lengths can be divided into three regions that correspond to what was observed in the separation energy curves. In the first (elastic) region, the bond lengths roughly follow a parabolic relationship as the separation distances increase for all three models. In the second (plastic) region, the changes of bond lengths are more drastic, and the different effects of B and O begin to become more obvious compared to the elastic region. For example, in Fig. 12(a-2), all of the interfacial Cu–Cu bonds exhibit an abrupt increase in length at a separation distance of 0.16 nm in the clean grain boundary. Combining this observation with the finding that the strong atomic interactions disappear immediately after a separation distance of 0.16 nm in Fig 11(a), it is clear that these bonds are broken at this separation distance. For the grain boundary with B at a CTP site, shown in Fig. 12(b-2), abrupt changes in bond lengths can be divided into two avalanche stages: (1) at a separation distance of 0.175 nm, where the Cu1–Cu2, Cu1–Cu5, Cu2–Cu2, Cu2–B and Cu3–B bonds break suddenly and (2) at a separation distance of 0.19 nm, where the Cu1–Cu3, Cu3–Cu4 and Cu4–Cu5 bonds break abruptly. The bond lengths of Cu1–Cu5, Cu3–Cu4 and Cu4–Cu5 bonds adjacent to the B atom are much shorter than the corresponding Cu–Cu bonds in the clean grain boundary during the first-principles-based tensile tests, which means that the B atom enhances the strengths of these bonds and increases the critical separation for fracture to 0.19 nm. For the grain boundary with O at a PBP site, almost all of the interfacial bonds break at the separation distance of 0.175 nm. Although the figure indicates that the O atom prolongs the critical separation for fracture up to 0.175 nm, it is important to recall that the O-doped boundary is still weaker than the clean grain boundary. In the third region, the bond lengths all increase linearly with increasing separation distance. With the detailed electronic and atomistic reconstructions during the straining, the opposite effects of B and O on the strength have been illuminated. Namely, B at the CTP site enhances the charge density among the closest Cu atoms to improve the connection of between the two grains, while O at the PBP site reduces these interactions between the Cu atoms.

## 4. Summary and Conclusions

In conclusion, the impact of common nonmetallic impurities on the energetics and mechanical strength of a Σ5 (310) grain boundary in Cu was investigated using first-principles simulations. The energetics analysis shows that all of the nonmetallic impurities increase thermodynamic stability by decreasing grain boundary energy. The grain boundary energy decreases as both the covalent radius of the impurities and the volumetric deformation of the structural polyhedra at the interface. Within a given period of the periodic table, the strengthening energy increases as the covalent radius of the impurities decreases and the electronegativity of the impurities increases. The local reconstruction of charge density with impurity doping also plays an important role in determining the grain boundary strength. However, this effect can be reversed and become detrimental to boundary strength if the impurity causes a large distortion of the local interfacial structure and a reduction in charge density. Herein, B strengthens the boundary, C has almost no effect while all other impurities act to weaken the grain boundary. First-principles-based tensile tests show that B at the CTP site increases the tensile strength of the grain boundary, while O segregating into the PBP site decreases this property. These results come from the fact that B at the CTP site enhances the charge density among the adjacent Cu atoms to improve the connection between the two grains, while O at the PBP site reduces these interactions between the Cu atoms. Our results show that nonmetallic impurities can have a wide range of effects, in some cases beneficial while in others harmful, on the thermodynamic stability and mechanical strength because of alterations at the atomistic and electronic levels.

**Acknowledgement:**

This research was supported by the U.S. Army Research Office under Grant W911NF-16-1-

0369, the National Natural Science Foundation of China (51472188 and 51521001), Natural Research Funds of Hubei Province (No. 2016CFB583), Fundamental Research Funds for the Central Universities in China and the "111" project (B13035).

Table 1 Parameters of fracture energetics from first-principles-based tensile test

| Simulation type | System | Fracture energy ($J/m^2$) | Ultimate strength ($J/m^2$) | Tensile strength (GPa) | Strengthening energy (eV) |
|---|---|---|---|---|---|
| Rigid | Clean boundary | 2.734 | | 19.057 | |
| | B at CTP site | 2.814 | | 19.268 | -0.417 |
| | O at PBP site | 2.520 | | 17.460 | 1.115 |
| Relaxed | Clean boundary | 2.266 | 1.869 | 14.959 | |
| | B at CTP site | 2.371 | 2.065 | 15.159 | -0.547 |
| | O at PBP site | 2.097 | 1.794 | 13.209 | 0.881 |

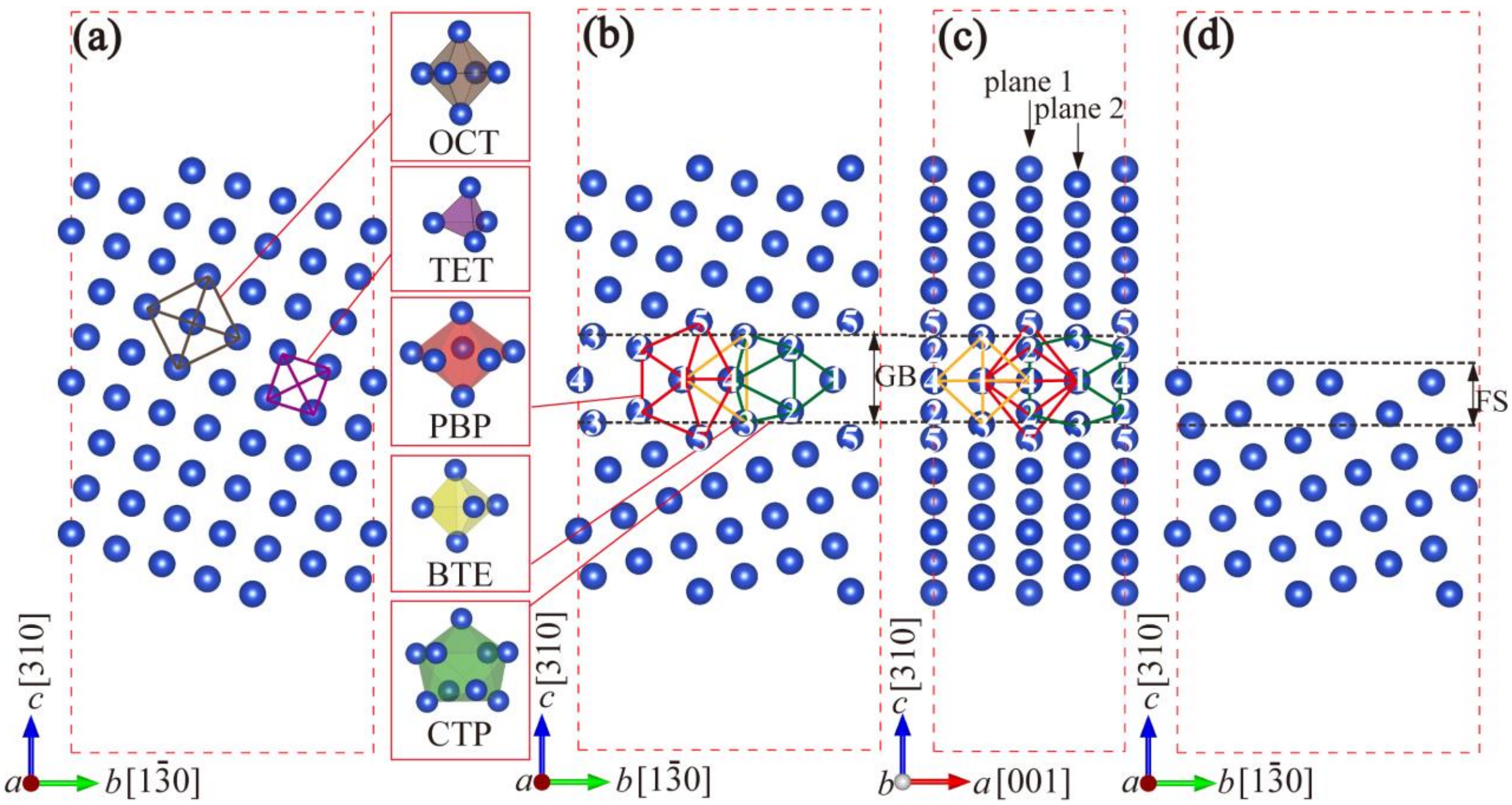

Fig. 1. Schematic illustration of (a) bulk Cu, (b) Σ5 (310) grain boundary (GB) viewed along the *a* axis, (c) Σ5 (310) grain boundary viewed along the *b* axis and (d) Cu (310) free surface (FS).

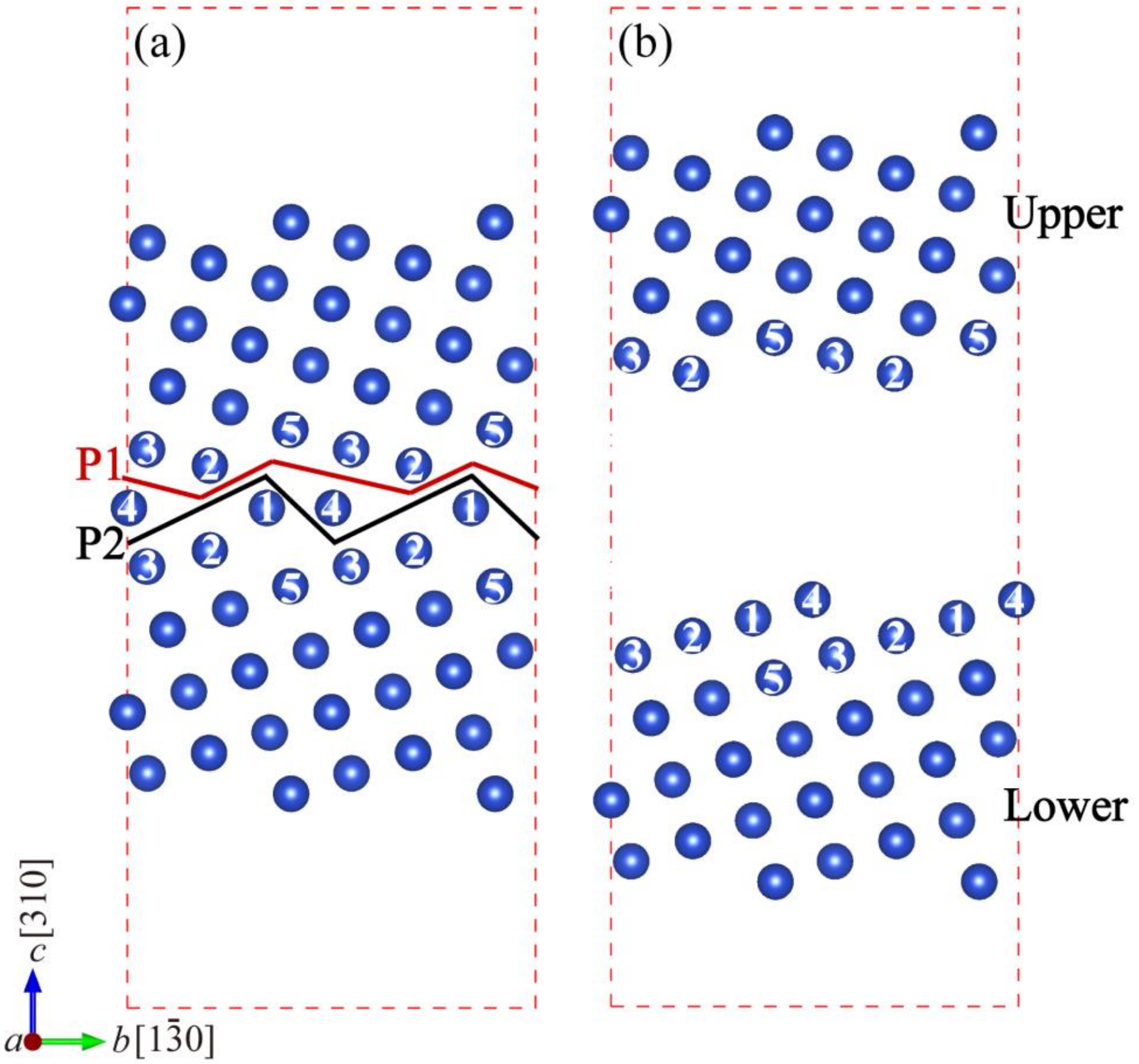

Fig. 2. Schematic diagram of (a) the fracture paths for the Cu grain boundary and (b) the totally separated two free surfaces. The P1 and P2 stand for the potential two fracture paths.

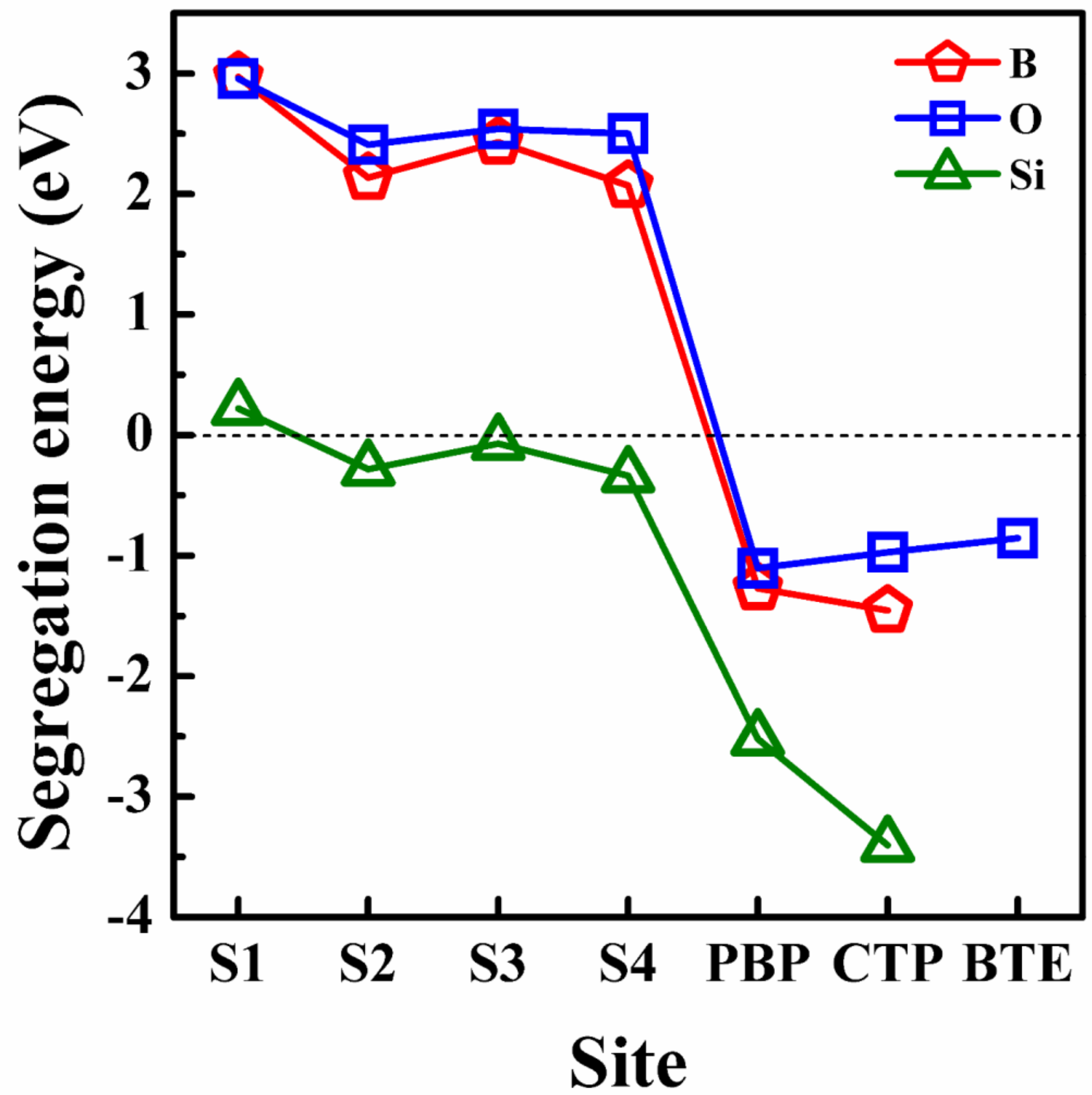

Fig. 3. Segregation energies of Cu Σ5 (310) grain boundaries with B, O and Si at different substitutional and interstitial sites. S1, S2, S3 and S4 stand for the substitutional sites of impurities at the interfaces.

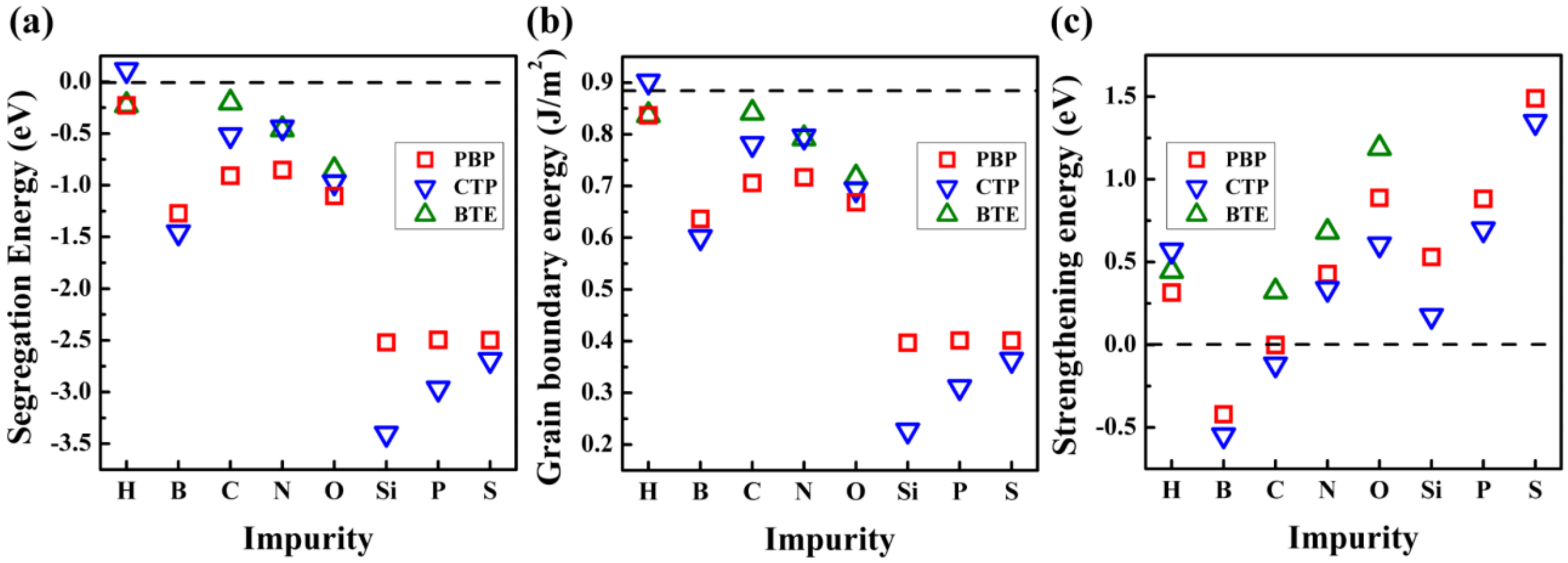

Fig. 4. (a) Segregation energies, (b) grain boundary energies and (c) strengthening energies of Cu grain boundaries with impurities in different polyhedra at the interfaces.

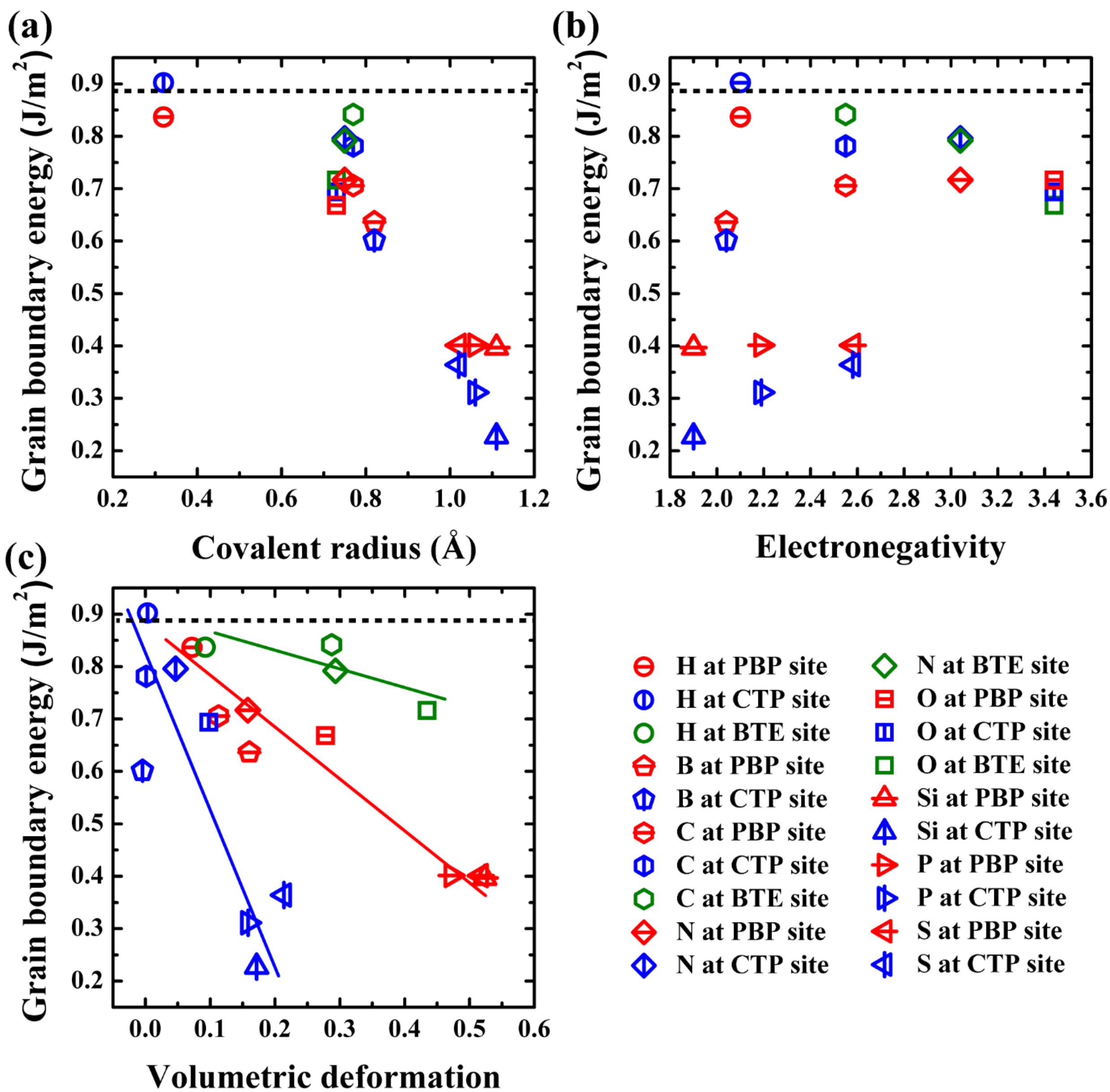

Fig. 5. Relationships between the grain boundary energy and (a) the covalent radius of impurities, (b) the electronegativity of impurities and (c) the volumetric deformation of polyhedra at the interface. The red, blue and green lines show the energetic change tendencies for the PBP, CTP and BTP sites with different impurities, respectively.

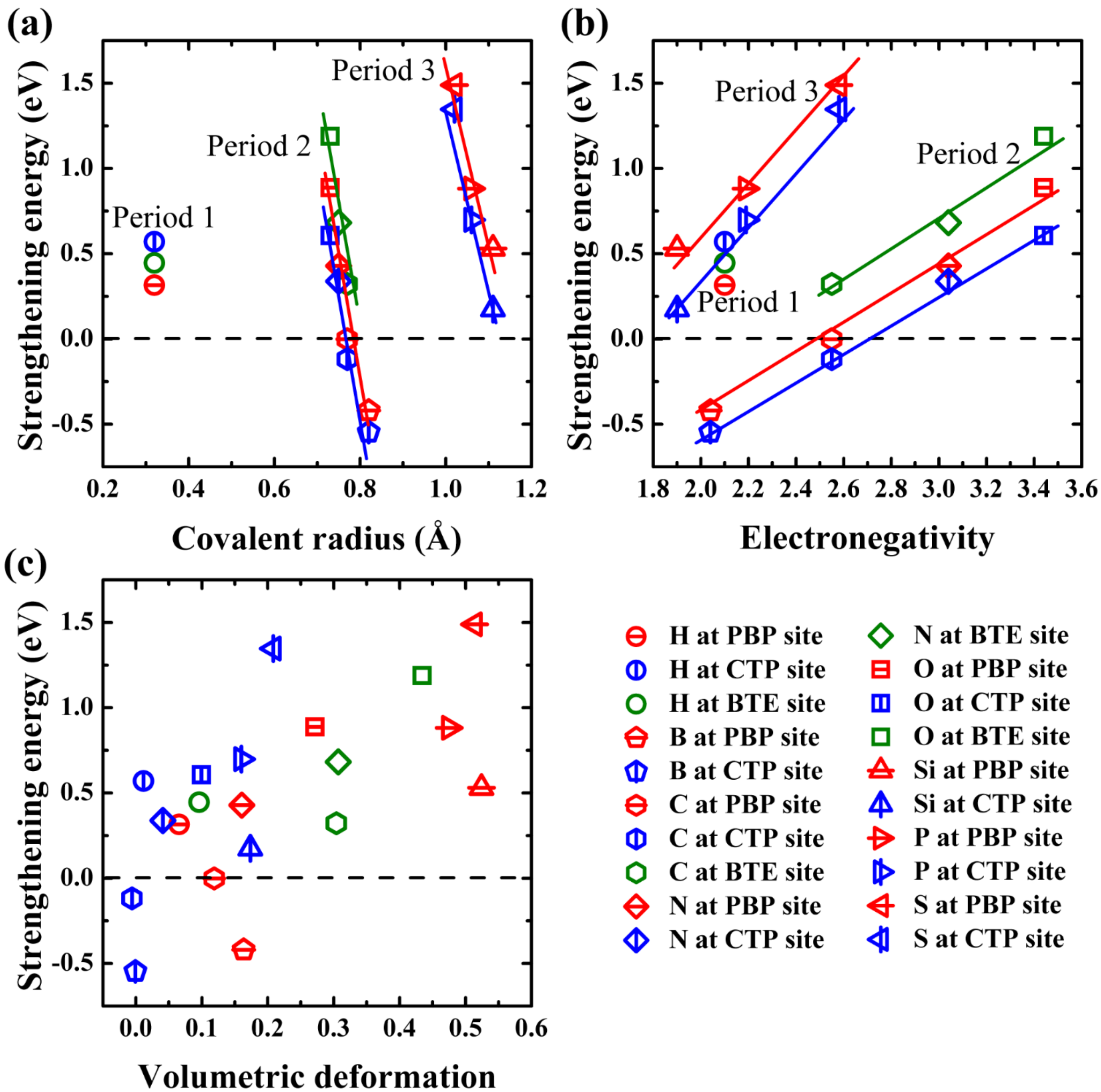

Fig. 6. Relationships between the strengthening energy and (a) the covalent radius of impurities, (b) the electronegativity of impurities, and (c) the volumetric deformation of polyhedra at the interface. The red, blue and green lines show the energetic change tendencies for the PBP, CTP and BTP sites with different impurities from different periods, respectively.

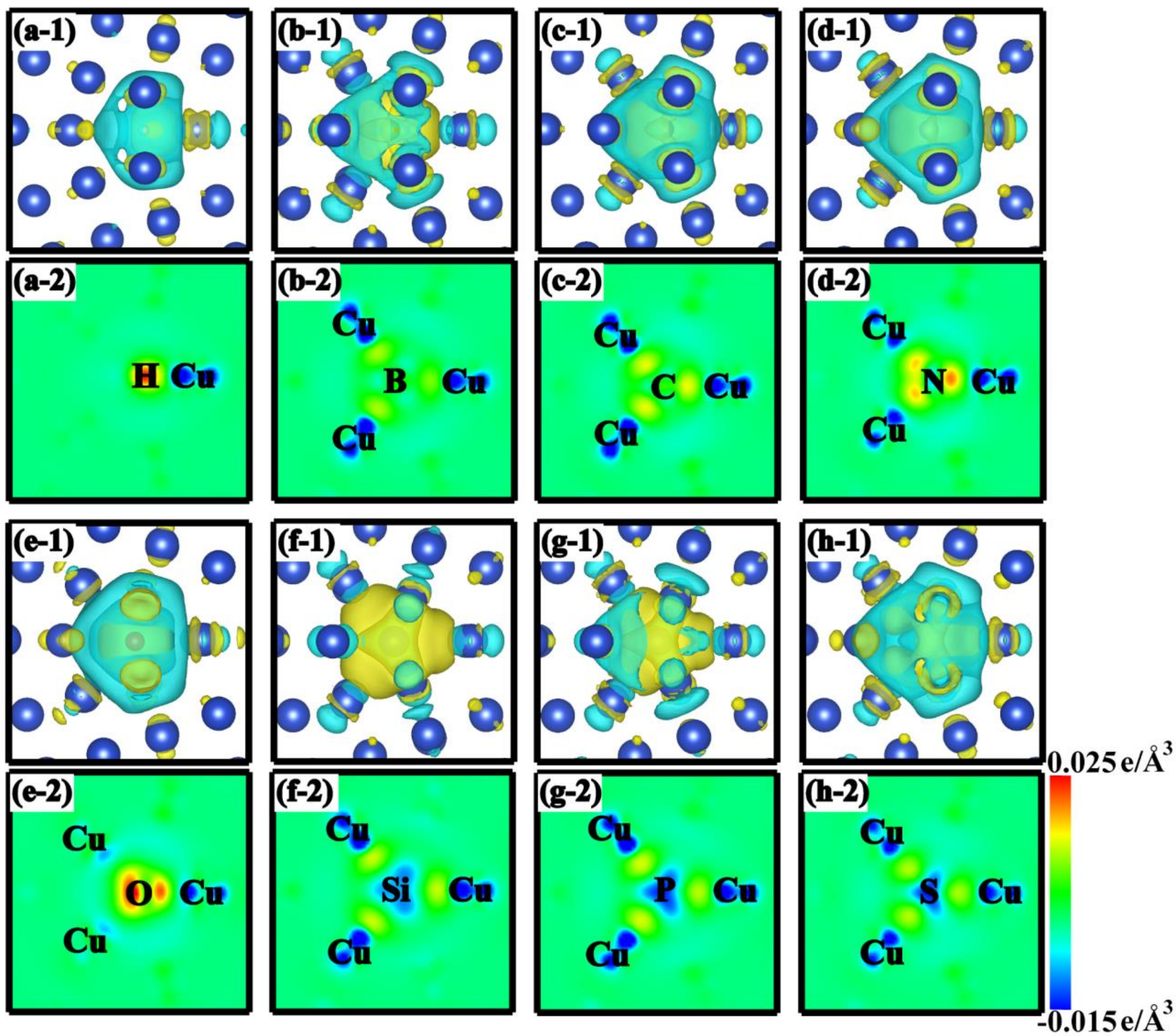

Fig. 7 Differential charge density of Cu Σ5 (310) grain boundaries with impurities: (a) H, (b) B, (c) C, (d) N, (e) O, (f) Si, (g) P and (h) S in the CTP site projected on the (100) plane. The number 1 denotes the three-dimensional view, while 2 denotes the two-dimensional charge density contours. In the three-dimensional charge density contours, the blue region represents electron depletion, while the yellow region signifies electron accumulation. In the two-dimensional charge density contours, the blue region represents electron depletion, and the red region signifies electron accumulation.

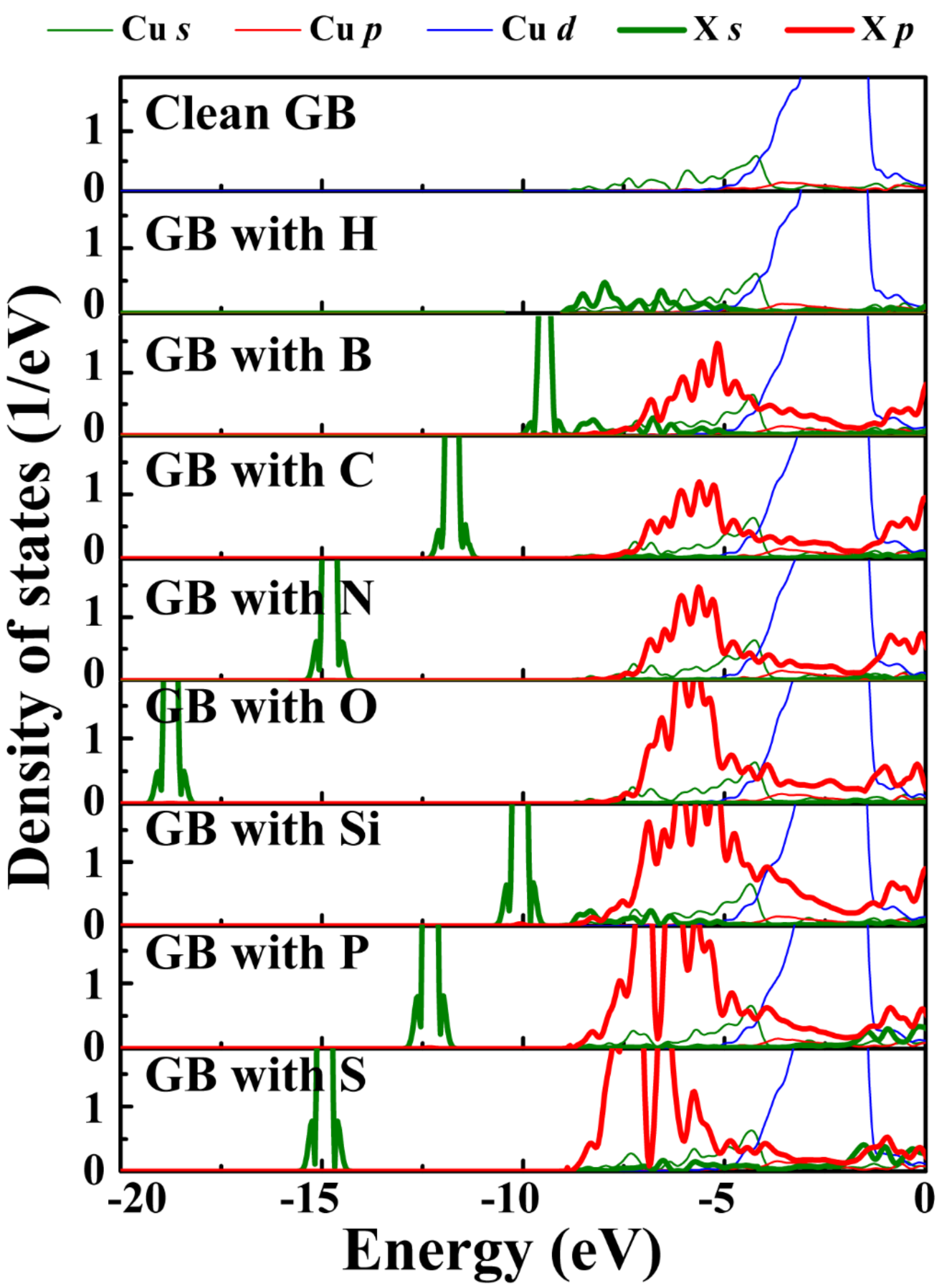

Fig. 8 Density of states for the impurity X and the closest Cu1 atom to X. The density of states data for Cu1 appears as thin lines, while the density of states data for the impurity X appears as thick lines.

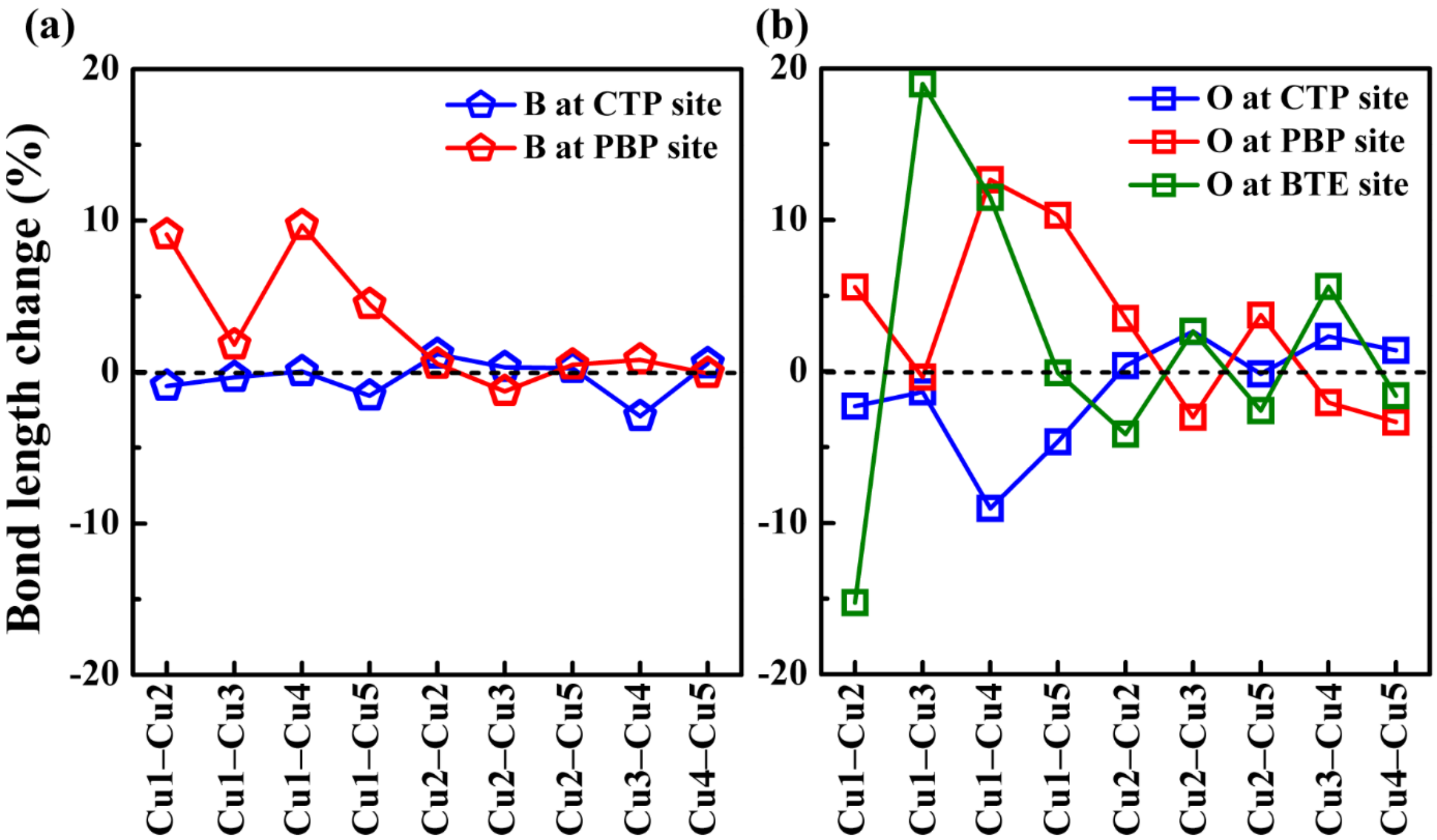

Fig. 9. Bond length changes of the Cu–Cu bonds closest to the impurity at the interfaces with (a) B and (b) O at different polyhedra sites.

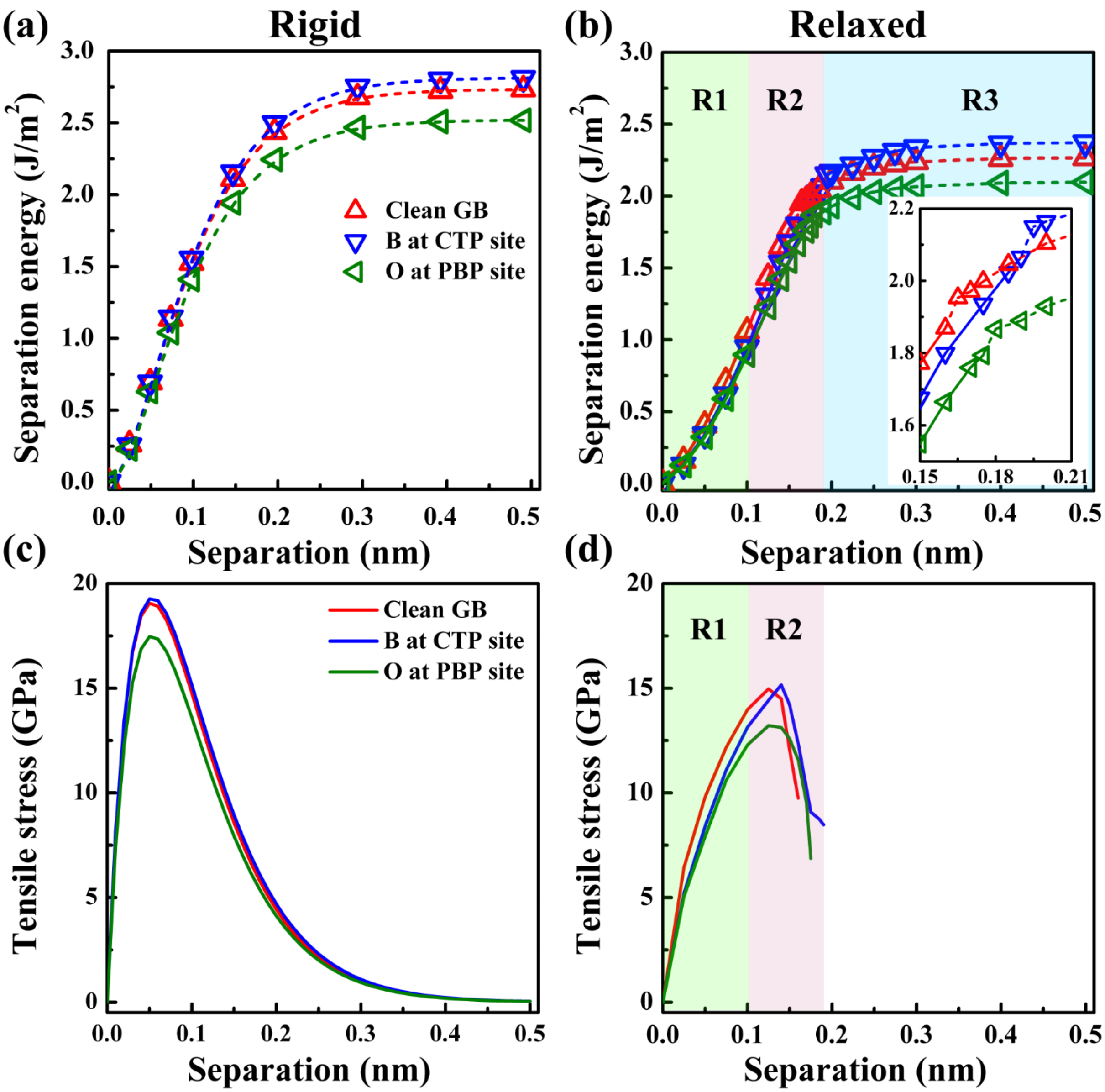

Fig. 10. The separation energy and tensile stress as a function of the separation distance for the three different samples. (a) Separation energies from the rigid calculations, (b) separation energies from the relaxed calculations, (c) tensile stresses from the rigid calculations, and (d) tensile stresses from the relaxed calculations. The dashed lines in (a) and (b) represent the fit curves by the universal binding energy relation proposed by Rose et al. [67]. The labels R1, R2 and R3 denote three distinctive regions during the first-principles-based tensile simulation.

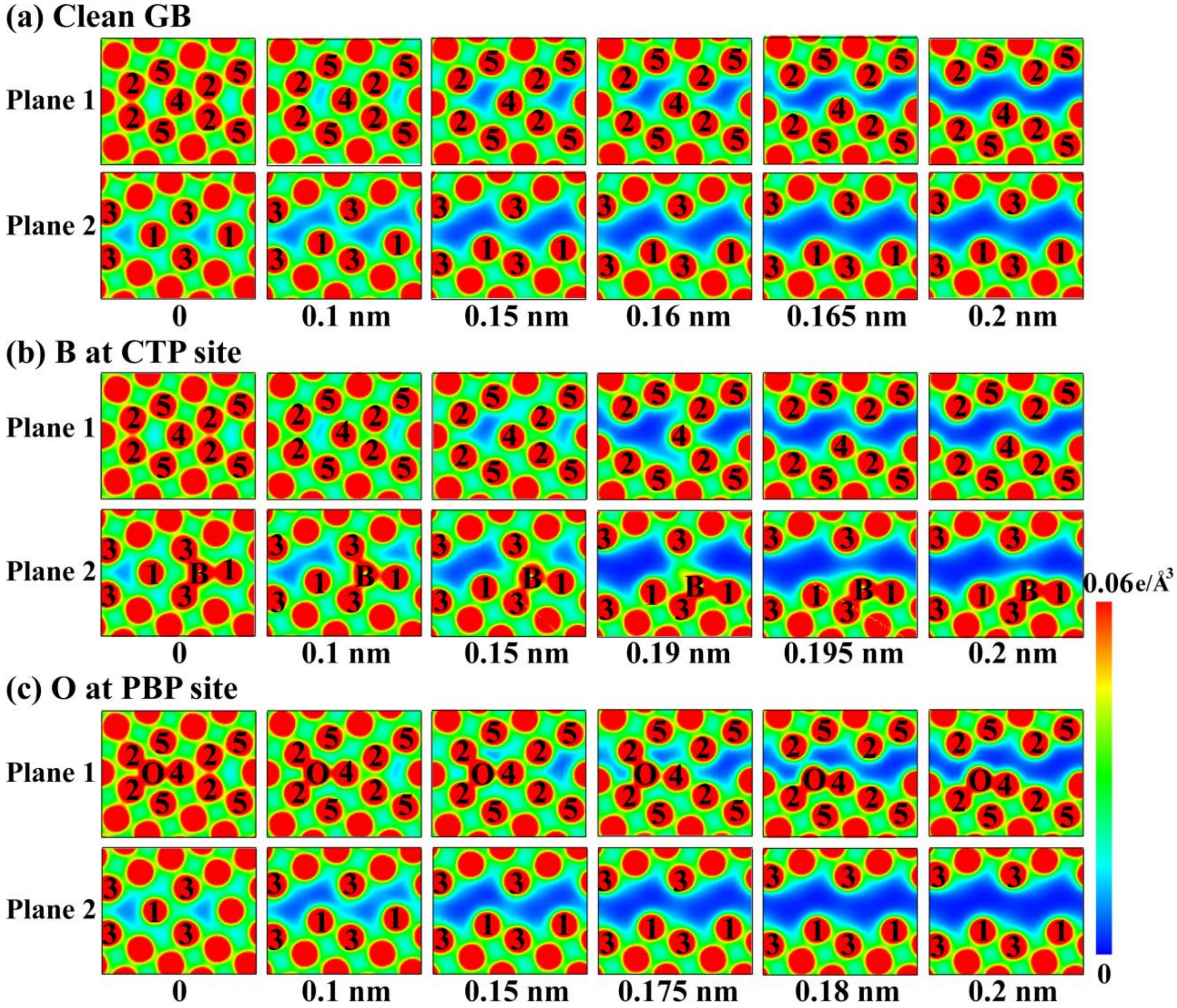

Fig. 11. Charge density distributions of (a) the clean Cu grain boundary and the grain boundary with (b) B at the CTP site and (c) O at the PBP site during relaxed first-principles-based tensile. Separation distance is shown below each frame.

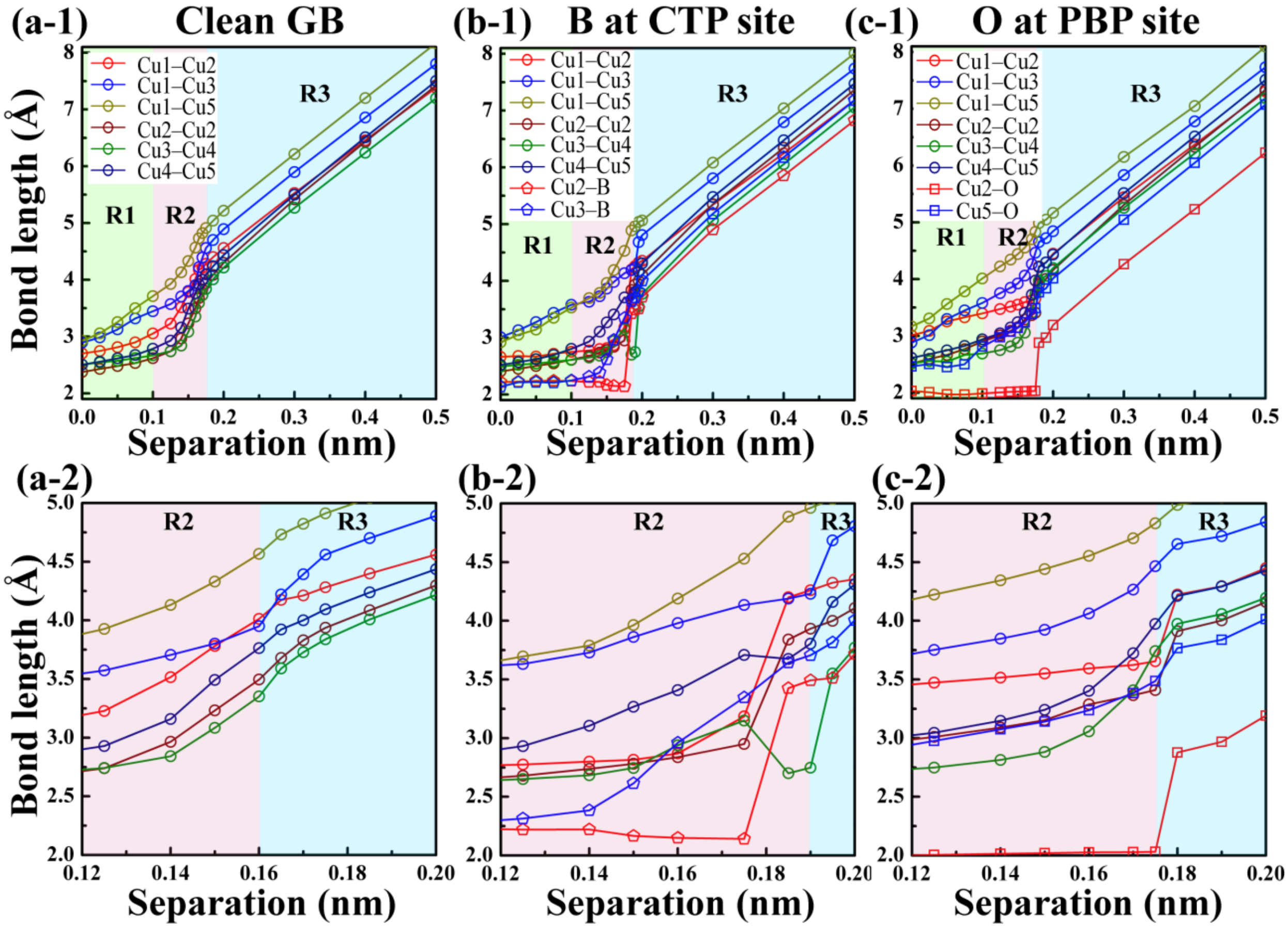

Fig. 12. The bond length of the interfacial bonds as a function of the separation distance of (a) the clean grain boundary, (b) the grain boundary with B at the CTP site and (c) the grain boundary with O at the PBP site. The figures marked as number 2 are the enlarged views of the figures marked as number 1. The labels R1, R2 and R3 denote the three distinctive regions during the first-principles-based tensile simulation.